\documentclass[a4paper,12pt]{JHEP3}
\usepackage{amsmath,amsfonts,epsfig}
\def\th#1.#2{\theta_1 (z_{#1} - z_{#2})}
\def\thn#1.#2.#3.#4{\theta_{\nu} (\frac{#1 z_{1} #2 z_{2} #3 z_3 #4 z_4}{2} )}
\def\us#1.#2.#3.#4{(u_{#1}\wedge u_{#2}) (u_{#3} \wedge u_{#4})} 
\def\bra{\langle}
\def\ket{\rangle}
\def\d{\mathrm{d}}
\def\bb#1{\mathbb{#1}}

\def\ap{\alpha^{\prime}}
\def\beq{\begin{equation}}
\def\eeq{\end{equation}}
\def\S{\tilde{S}}

\numberwithin{equation}{section}

\preprint{IPPP/06/92\\ DCPT/06/184\\hep-th/0612110}

\author{Steven A. Abel and Mark D. Goodsell\\ Centre for Particle Theory, University of Durham, Science Laboratories, South Road, Durham, DH1 3LE, UK\\
 E-mail: \email{s.a.abel@durham.ac.uk}, \email{m.d.goodsell@durham.ac.uk}}

\title{Realistic Yukawa Couplings through Instantons in Intersecting Brane Worlds}

\abstract{The Yukawa couplings of the simpler models of D-branes on 
toroidal orientifolds suffer from the so-called ``rank one'' problem -- there 
is only a single non-zero mass and no mixing. We consider the one-loop contribution 
of E2-instantons to Yukawa couplings on intersecting D6-branes, and show 
that they can solve the rank one problem.
In addition they have the potential to provide a geometric explanation 
for the hierarchies observed in the Yukawa coupling. In order to 
do this we provide the necessary quantities for instanton calculus in 
this class of background.}

\begin{document}

\section{Introduction}

Intersecting D-branes are an interesting possibility for string model building, 
allowing one to build models that come remarkably close to the MSSM in terms
of particle content and gauge group. 
A particularly simple and calculable subset of these are models constructed from 
 D6-branes in type IIA string theory, wrapping orientifolds of 
$\bb{R}_4 \times \bb{T}_2 \times \bb{T}_2 \times \bb{T}_2$ (see
e.g. \cite{Blumenhagen:1999ev}-\cite{Chen:2006sd}
or \cite{Blumenhagen:2005mu} for a review). One of the interesting
features of these models is that the localization
of matter fields at D-brane intersections may have implications for a
number of phenomenological questions, most notably the flavour 
structure of the Yukawa couplings \cite{Cremades:yukawa}. Initially it was thought 
that these could be understood by
having the matter fields in the coupling located at different
intersections, with the resulting coupling being suppressed by the
classical world-sheet instanton action (the minimal world-sheet area
in other words). Since the resulting Yukawa couplings are of the form  
\[ 
Y \propto e^{-Area/2\pi\alpha'}
\] 
the hope was that one would be able to find a
geometric explanation of the mass and mixing hierarchies of the MSSM. 
Not only would this be 
simple and elegant, but it would also mean that measurements of 
the Yukawa couplings of the MSSM would yield direct information 
about the compactification geometry. Unfortunately in the simplest 
compactifications this hope was misplaced. The reason why 
is the so-called ``rank one'' problem; the simplest models 
have left-handed fields separated in one $\bb{T}_2$, and right-handed 
fields separated in a different $\bb{T}_2$, and the couplings 
had a rank one flavour structure, 
\[ 
Y^0_{ab} \sim A_a B_b,   
\] 
where $a,b$ label flavour and $A,B$ are two vectors whose elements are
sensitive to the displacements of vertices in the compact dimensions. 
Only the third generation gets a mass and there is no mixing. 

There have been numerous subsequent attempts to solve this problem 
as well as other related analyses of questions regarding flavour 
(e.g. refs.\cite{Abel:2003fk}-\cite{Dutta:2006bp}), but many of these 
lost the original link with geometry. In parallel there developed 
techniques for calculating both  
tree level \cite{Cvetic:Yukawa}-\cite{Bertolini:2005qh}, and 
loop \cite{Abel:2004ue}-\cite{Abel:2005qn} amplitudes involving
chiral (intersection) states on networks of intersecting D-branes. 

The most recent development on the calculational side, 
whose consequences will be the subject of this paper, has been the 
incorporation of the effect of instantons in ref.\cite{Blumenhagen:2006xt} 
(for related work see also 
\cite{Billo:2002hm}-\cite{Florea:2006si}). This 
work laid out in detail how the contributions of so-called E2-instantons 
(i.e. branes with 3 Neumann boundary conditions in the compact dimensions 
and Dirichlet boundary conditions everywhere else) to the superpotential 
could be calculated. It also pointed out a number of expected 
phenomenological consequences of these 
objects. Thus far attention has mostly been paid to the fact that instantons 
do not necessarily respect the global symmetries of the effective 
theory and so are able to generate terms that may otherwise be disallowed. In particular 
they can be charged under the parent anomalous U(1)'s due to the Green-Schwarz anomaly 
cancellation mechanism. (Alternatively, these charges can be associated with states at the 
intersection of the E2 and D6 branes.) For 
example this can induce Majorana mass terms for neutrinos which are of the form 
\[ 
e^{-\frac{8\pi^2}{g_{E2}^2}} M_{s} \nu_R \nu_R\, ,
\] 
and which would otherwise be forbidden. 
In this equation $g_{E2}$ is an effective coupling strength which depends on the 
world-volume of the E2 instanton. These need not be equal to the gauge couplings 
of the MSSM and the Majorana mass-terms can be of the right order to generate 
the observed neutrino masses \cite{Blumenhagen:2006xt},\cite{Ibanez:2006da}. Similarly 
interesting contributions occur for the $\mu$-term of the MSSM, yielding 
a possible solution to the $\mu$-problem. 

In this paper we reassess Yukawa couplings in the light of 
instanton contributions. In particular we claim that one-loop 
diagrams with E2 branes can solve the rank-one problem and lead to a 
Yukawa structure which is hierarchical. 
 All the Yukawa couplings have by 
assumption the same charges under all symmetries, so the extra 
terms are induced by E2 instantons which do not intersect the D6-branes.
The tree and one-loop contributions to Yukawa couplings are shown 
in figure 1; the tree level diagram consists of the usual disc diagram
with three vertices, the 
one loop diagram is an annulus with the inside boundary being an E2 brane and three
vertices on the D6-brane boundary.
By explicit computation of these one-loop instanton contributions 
we show that the corrected Yukawas are of the form 
\[ 
Y_{ab}=Y^0_{ab}+ Y^{np}_{ab}, 
\] 
where the nonperturbative contribution has rank 3. 
Moreover as for the neutrino Majorana masses, these contributions are 
exponentially suppressed by the instanton volume. Hence the 1st and 2nd generation
masses are hierarchically smaller than the first. 

In addition there is the possibility of making
an interesting connection between the Yukawa hierarchies and the Majorana 
neutrino masses. The suppression of the latter with respect to the 
string scale should be similar to the suppression of 2nd generation 
masses with respect to third generation ones. In general one sees 
that a direct connection between compactification geometry and 
Yukawa couplings would be manifest. 
The rest of this paper is devoted to proving this result. In the 
next section we outline the techniques of instanton 
calculus and compute the necessary general results: these include the 
multiplying factors of disconnected tree-level and one-loop diagrams 
with no vertex operators, which must appear in every such process. 
The section that follows provides the annulus correlator, and in particular 
shows that the leading contribution yields instanton contributions to
the Yukawa couplings of the advertised form.  

\begin{figure}
\begin{center}
\epsfig{file=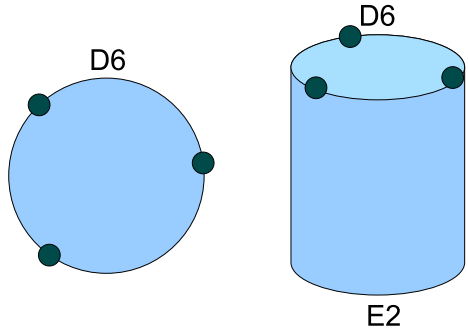,height=5.0in}
\vspace{-2.5in}
\caption{Contributions to 3 point functions: tree level and one-loop instanton.}\label{fig:annulus}\end{center}
\end{figure}

\section{Instanton Calculus}
\label{Calculus}

The general framework for calculating E2 instanton corrections to the
superpotential in string backgrounds was proposed in
\cite{Blumenhagen:2006xt}. This section elucidates the technical
details of the calculus of E2 instanton corrections applied to
toroidal orientifold intersecting brane worlds.

\subsection{Tree Level Contributions}

The non-perturbative $O(e^{-8\pi^2/g_{E2}^2})$ factor appearing in
instanton contributions to the superpotential is given by the product
of all possible disk diagrams whose boundary lies along the instanton
and with either no vertex operators or an RR-tadpole operator. We
shall label this contribution $e^{-S^0_{E2}}$. It is given by
\beq
S^0_{E2} = \frac{2\pi}{g_s \sqrt{(\ap)^3}} {L_{E2}}
\eeq
where $L_{E2} = \prod_{\kappa=1}^3 L^{\kappa}_{E2}$ is the volume of the instanton. The above can be given more conveniently in terms of the gauge coupling on a reference brane $a$ as 
\beq
S^0_{E2} = 8\pi^2 \frac{L_{E2}}{L_a} g_a^{-2} 
\eeq  

\subsection{Pfaffians and Determinants}

To perform any calculation in an E2 instanton background requires the knowledge of the reduced Pfaffian and Determinant factor given by the exponential of the total partition function of states intersecting the instanton with the zero modes removed. In toroidal backgrounds, there are two classes of contributions to this: the first arises when the instanton is parallel to a $D6$-brane in one torus, and the second appears when there is no parallel direction. 

\subsubsection{One Parallel Direction}

In this case, there is no zero mode associated provided that the branes do not intersect, i.e. brane $a$ is separated from the E2 brane by some distance $y_{a,E2}$ in a torus $i$. In this case the partition function is given by

\begin{eqnarray}
Z(E2, D6_a) & = &  N_a I_{E2,a}^{j} 
I_{E2,a}^{k}\int \frac{\d t}{t} 
\sum_{\nu} \sum_{r^{i},s^{i}} \delta_{\nu} \frac{\theta_{\nu}^2 (\frac{it}{2})}{\theta_{1}^2 (\frac{it}{2})} \frac{\theta_{\nu} (\phi_{E2,a} i t) \theta_{\nu} (-\phi_{E2,a} i t)}{\theta_{1} (\phi_{E2,a} i t) \theta_{1} (-\phi_{E2,a} i t)} \nonumber \\
&& \, \, \, \qquad \qquad \qquad \times
\exp \bigg[ \frac{-8\pi^3 \alpha^{\prime} t}{(L_{E2}^{i})^2} |r^{i} + \frac{iT_2^{i} s^{i}}{\alpha^{\prime}} + \frac{i y_{a,E2} L_{E2}^{i}}{4\pi^2 \ap}|^2 \bigg] 
\end{eqnarray}
After performing the sum over spin structures we obtain an expression that would 
also be found in gauge threshold correction calculations
\beq
Z(E2, D6_a) = N_a I_{E2,a}^{j} I_{E2,a}^{k} \int \frac{\d t}{t} \sum_{r^{i},s^{i}} \exp \bigg[ \frac{-8\pi^3 \alpha^{\prime} t}{(L_{E2}^{i})^2} |r^{i} + \frac{iT_2^{i} s^{i}}{\alpha^{\prime}} + \frac{i y_{a,E2} L_{E2}^{i}}{4\pi^2 \ap}|^2 \bigg] .
\eeq
This expression can be integrated. When all $y_{a,E2}\rightarrow 0$ (where there are fermionic zero modes that must be regulated) it gives \cite{Lust:Gauge}
\beq
Z(E2, D6_a)_{y=0} = -N_a I_{E2,a}^{j} I_{E2,a}^{k} \left[ \ln \frac{(L_{E2}^i)^2}{\ap} + \ln | \eta(\frac{T^i}{\ap}) |^4 + \kappa \right ],
\eeq
where $\kappa = \gamma_E - \log 4\pi$ is a moduli-independent constant. For non-zero $y_{a,E2}$,  the brane separation serves as an IR regulator to give 
\beq
Z(E2, D6_a) = 
- N_a I_{E2,a}^{j} I_{E2,a}^{k} \bigg[\log \left| \frac{\theta_1 (\frac{i y_{a, E2} L_{E2}^i}{2\pi^2 \ap}, 
\frac{i T_2^i}{\ap})}{\eta(\frac{iT_2^i}{\ap})} \right|^2 - \frac{y_{a, E2}^2}{2\pi^3 \ap} 
\frac{(L_{E2}^i)^2}{T_2^i}\bigg]\, .
\eeq
This is derived in appendix \ref{N2Sector} and agrees with \cite{Berg:2004ek}\footnote{The authors are grateful to the referee for mentioning this reference.}

It is noteworthy that the IR regularisation by separation of the $E2$ brane from brane $a$ does not commute with that used for $y_{a, E2}=0$. 
This reflects the qualitative difference between E2 branes which 
wrap cycles where $b_1 (\Pi_{E2})=0$ and those where the betti number is non-zero; 
in the latter case (i.e. when
the instanton does not pass through a fixed point) it acquires
additional uncharged fermionic zero modes which must be integrated
over, but when the separation between D6 and E2 branes reduces to zero
there arise additional charged fermionic zero modes. We only consider
configurations with no additional zero modes in this paper, and thus
we require the E2 branes to pass through fixed points and D6 branes to
be separated from these by some small distances. This is a common
feature of many models, although it is incompatible with the models
of, for example, \cite{Blumenhagen:2005tn,Chen:2006sd}. However, since
the extra uncharged modes carry no coupling, it is actually
possible for E2 instantons with six additional uncharged zero modes to
contribute to the superpotential term considered in this paper
generated at one loop, and thus we expect our conclusions to extend to
those models as well. 
As explored recently in \cite{Bianchi:2007fx,Akerblom:2007uc,Blumenhagen:2007bn}, this can occur when there are non-zero bulk
fluxes threading the E2 brane worldvolume (which are naturally present in models
to stabilise moduli), but in that case the exact calculation of 
the superpotential contribution is beyond the scope of this paper.

\subsubsection{No Parallel Directions}

Here we have the raw expression
\begin{equation}
Z(E2, D6_a) = N_a I_{E2,a} \int \frac{\d t}{t} \sum_{\nu}  \delta_{\nu} \frac{\theta_{\nu}^2 (\frac{it}{2})}{\theta_{1}^2 (\frac{it}{2})} \frac{\eta^3 (it)}{\theta_{\nu} (0)} \prod_{\kappa=1}^3 \frac{\theta_\nu (\phi^{\kappa}_{E2 ,a} i t)}{\theta_1 (\phi^{\kappa}_{E2,a} i t)}.
\end{equation}
However, this amplitude is not defined in the $\nu=1$ sector, and
additionally contains fermionic zero modes which must be
regularised. The first issue is straightforward to resolve: provided
that the model satisfies RR-charge cancellation, the $\nu=1$ sector
does not contribute. To address the second issue we must decide how to
remove the zero mode from the trace over states: this is trivial once
we express the amplitude in the open string channel. Since the
integral over the modular parameter and the exponentiation of the
partition function translates the sum over states into a product, it
is clear that we must simply expand the partition function as a series
in $e^{-\pi t}$ and remove the $O(1)$ term in the $\nu=1,2$
sectors. The piece subtracted turns out to be cancelled over all the
contributions by the RR cancellation condition, and so we can ignore
it. 

We must now perform the sum over spin structures. To this end, we use the identity
\begin{equation}
\left( \frac{\theta_\nu (z) \theta_1^{\prime} (0)}{\theta_1 (z) \theta_{\nu}(0)} \right)^{2} = \frac{\theta_{\nu}^{\prime \prime} (0)}{\theta_{\nu} (0)} - \partial^2 \log \theta_1 (z)
\end{equation}
to write the partition function as
\begin{equation}
Z(E2, D6_a) = N_a I_{E2,a} \int \frac{\d t}{t} \sum_{\nu \neq 1}  \delta_{\nu} \bigg[ \theta_{\nu}^{\prime \prime} (0) - \theta_{\nu} (0)\partial^2 \log \theta_1 (\frac{it}{2}) \bigg] \frac{1}{2\pi \theta_1^{\prime} (0)} \prod_{\kappa=1}^3 \frac{\theta_\nu (\phi^{\kappa}_{E2,a} i t)}{\theta_1 (\phi^{\kappa}_{E2,a} i t)}.
\end{equation}
It is straightforward to show that, if we were to include the $\nu=1$ term to the above, it would not contribute, and hence we can perform the spin structure sum using the usual Riemann theta identities. The term proportional to $\theta_{\nu}(0)$ vanishes, and we have remaining an expression identical to that found in gauge threshold corrections. The calculation is then that of \cite{Lust:Gauge,Akerblom:2007np}, including the cancellation of divergences arising in the NS sector. The result is then

\begin{multline}
\frac{\mathrm{Pfaff}^{\prime} (D_F)}{\sqrt{\det^{\prime} (D_B)}} = \prod_{k} \left(\frac{\Gamma(1-2\phi_{E2,k}^{1})\Gamma(1-2\phi_{E2,k}^{2})\Gamma(1+2\phi_{E2,k}^{1}+2\phi_{E2,k}^{2})}{\Gamma(1+2\phi_{E2,k}^{1})\Gamma(1+2\phi_{E2,k}^{2})\Gamma(1-2\phi_{E2,k}^{1}-2\phi_{E2,k}^{2})}\right)^{-4I_{E2,O6_k}}\\
\prod_{a=\{a,a^{\prime}\}} \left(\frac{\Gamma(1-\phi_{E2,a}^{1})\Gamma(1-\phi_{E2,a}^{2})\Gamma(1+\phi_{E2,a}^{1}+\phi_{E2,a}^{2})}{\Gamma(1+\phi_{E2,a}^{1})\Gamma(1+\phi_{E2,a}^{2})\Gamma(1-\phi_{E2,a}^{1}-\phi_{E2,a}^{2})}\right)^{N_a I_{E2,a}}
\end{multline}

This result, in contrast to that of the previous subsection, is not a holomorphic function of the moduli fields, and hence cannot contribute to the superpotential. This issue has recently been explored in \cite{Akerblom:2007uc}, where they determined that the above generates the supergravity factor $e^{\frac{K}{2}}$ ($K$ being the K\"ahler potential) and K\"ahler potential normalisation for the charged zero modes, so indeed no contribution to the superpotential.

\subsection{Vertex Operators and Zero Modes}

Massless strings with at least one end on the E2 brane are the zero modes of the instanton. The most important of these (in that they are always present) are the fermionic modes with both ends on the E2 brane; these are the modes associated with the two broken supersymmetries in the spacetime dimensions. The vertex operator for toroidal models is thus
\beq
V^{-1/2}_{\theta} = \theta_i \S^i \ e^{-\frac{\phi}{2}} \ \theta^{I}_{\alpha} \Theta^{\alpha}
\eeq 
where the internal spin field is $ \Theta^{\alpha} = \prod_{\kappa=1}^3 e^{ \pm \frac{ i}{2} H^{\kappa}}$; it is the spectral flow operator for the internal dimensions, or alternatively the internal part of the supercharge. This generically has four components after the GSO projection, and thus we have the eight modes of (broken) $N=4$, but only the two modes that preserve the same $N=1$ supersymetry as the branes and orientifolds in the model will contribute. In practice this involves choosing the correct $H$-charges to transform a fermion into a boson in the internal dimensions, and without loss of generality we shall take this to be $\prod_{\kappa=1}^3 e^{- \frac{ i}{2} H^{\kappa}}$. In addition, we may be concerned about antichiral bosonic zero modes that would spoil the generation of a superpotential. However, as discussed in \cite{Bianchi:2007wy,Argurio:2007vqa,Blumenhagen:2007bn}, provided that the E2 brane is invariant under the orientifold projection, these will be removed from the spectrum.

Massless fermions at an intersection between the E2 brane and a D6 brane of the model are internal zero modes of the instanton. They will not be relevant for the following analysis, but are in general vitally important for calculations; we list their vertex operators here for completeness:
\begin{eqnarray}
V_{\lambda_a} &=& \lambda_{a,I}^i e^{-\frac{\phi}{2}} \Delta \prod_{\kappa=1}^3 e^{i (\phi_{a E2}^{\kappa} - 1/2)H^{\kappa}} \sigma_{\phi_{a E2}^{\kappa}} \nonumber \\
V_{\bar{\lambda_b}} &=& \bar{\lambda}_{b,I^\prime}^i e^{-\frac{\phi}{2}} \bar{\Delta} \prod_{\kappa=1}^3 e^{i (\phi_{E2 b}^{\kappa} - 1/2)H^{\kappa}} \sigma_{\phi_{E2 b}^{\kappa}}
\end{eqnarray}
where $I$ is the intersection number running from $1,...,[\Pi_{E2} \cap \Pi_a]^+$, and $I^{\prime}$ runs from  $1,...,[\Pi_{E2} \cap \Pi_b]^-$ (note that $\phi_{E2 b} - 1/2 = -(\phi_{b E2}-1/2)$); $i=1,..,N_a (N_b)$ is the Chan-Paton index. $\Delta$ and $\bar{\Delta}$ are the boundary-changing operators in the four non-compact dimensions which interpolate between Dirichlet and Neumann boundary conditions; their OPE is
\beq
\Delta (z) \bar{\Delta}(w) \sim (z-w)^{-1/2} \, .
\eeq

\section{Yukawa Coupling Corrections}
\label{YukCup}

A straightforward analysis of the possible diagrams in the instanton calculus shows that in the presence of fermionic zero modes an $E2$-instanton cannot contribute to a Yukawa term in the superpotential for the quarks. However, if there exists one or more sLag cycles for which there are no intersections with any branes in the model, then a contribution to this term is possible through a one-loop annulus diagram. Typically this involves the separation of the $E2$-instanton from each D6 brane in one subtorus only, so that were the $E2$ brane replaced by a D6 brane wrapping the same three-cycle and the separations reduced to zero, this brane would preserve different $N=2$ supersymmetries with each of the branes in the model. This possibility arises generically but not in every case in model-building, so implies a new moderate constraint, in order 
to benefit from the consequences of these instantons.

To summarise the above and the conclusions of the previous section, in order to generate a correction to allowed superpotential couplings, then, we require
\begin{itemize}
\item $b_1(\Pi_{E2}) = 0$ 
\item $[\Pi_{E2} \cap \Pi_a] = 0 \qquad \forall a$ 
\item E2 brane invariant under orientifold projection, \emph{or} suitable fluxes lifting bosonic zero modes
\end{itemize}
In the following we shall determine a contribution to the superpotential for an orientifold of $\bb{R}_4 \times \bb{T}_2 \times \bb{T}_2 \times \bb{T}_2$, but expect our conclusions more generally to any model satisfying the above conditions, the further exploration of which we postpone to future work. In particular, we shall consider fractional E2 branes (under a suitable orbifold projection) to satisfy $b_1(\Pi_{E2}) = 0$ (for example in models with discrete torsion such as \cite{Blumenhagen:2005tn,Chen:2006sd,Cvetic:2007ku}) but bulk D6 branes. The annulus will only couple to the bulk component of the E2 brane, and hence the calculation is insensitive to the details of the orbifold projection.

A Yukawa superpotential term generated by E2 instantons
involves three superfields on an annulus diagram, and thus two fermions and one boson together with two fermionic supersymmetry zero modes. We can then write (with vertex operators in the field theory basis, normalised by $V_{\phi_{ab}} \rightarrow \sqrt{K_{\phi_{ab} \bar{\phi}_{ab}}} V_{\phi_{ab}}$ from the string basis)\cite{Akerblom:2007uc}:
\beq
e^{\frac{K}{2}} W_{np} = \int \d^4 x \d^2 \theta \langle V^{0}_{\phi_{ab}} V^{1/2}_{\psi_{bc}} V^{1/2}_{\psi_{ca}} V^{-1/2}_{\theta} V^{-1/2}_{\theta}\rangle_{a,E2} \frac{\mathrm{Pfaff}^{\prime} (D_F)}{\sqrt{\det^{\prime} (D_B)}} e^{-S_{E2}^0}.
\eeq

Since there are now three picture-changing operators in the above amplitude, to obtain a non-zero result we must apply each operator to a different subtorus direction; this is since each internal fermionic correlator must have zero net charge, and thus the charges introduced by the supersymmetry zero modes must be cancelled by those of the PCOs. The amplitude will then have no momentum prefactors, and thus will not factorise onto a scalar propagator; this is explicitly shown in appendix \ref{SpinSum}. 

Having determined the Pfaffian and Tree-level factors in section \ref{Calculus}, we must now determine the annulus correlator. The total amplitude can be written as
\begin{multline}
\langle V^{0}_{\phi_{ab}} V^{1/2}_{\psi_{bc}} V^{1/2}_{\psi_{ca}} V^{-1/2}_{\theta} V^{-1/2}_{\theta}\rangle_{a,E2} = \phi_{(ab)} \psi_{(bc)\alpha} C^{\alpha \beta} \psi_{(ca)\beta}\  \theta_1 \theta_2  \\
\times \int \d t \ it \prod_{i=2}^3 \int_0^{it} \d z_i \prod_{j=1}^2 \int_{1/2}^{1/2+it} \d w_j \langle e^{\frac{\phi}{2}(z_2)} e^{\frac{\phi}{2}(z_3)} e^{-\frac{\phi}{2}(w_1)} e^{-\frac{\phi}{2}(w_2)} \ket \lim_{x_1 \rightarrow z_1} \lim_{x_2\rightarrow z_2} \lim_{x_3 \rightarrow z_3}\\
\times (x_1-z_1) (x_2-z_2)^{1/2} (x_3-z_3)^{1/2}
\bra \S^{1}(z_2) \S^1 (z_3) \S^2 (w_1) \S^{2} (w_2)\ket \bra \prod_{i=1}^3 e^{i k_i \cdot X (z_i)} \ket  \\
\sum_{\{y_1,y_2,y_3\} = P(x_1,x_2,x_3)} \prod_{\kappa=1}^3 \sqrt{\frac{2}{\ap}}\langle \partial \bar{X}^{\kappa} (y_\kappa) \sigma_{\phi^{\kappa}_{ab}} (z_1) \sigma_{\phi^{\kappa}_{bc}} (z_2)\sigma_{\phi^{\kappa}_{ca}}(z_3) \ket \\
\times\bra e^{i H^{\kappa} (y_{\kappa})} e^{i(\phi_{(ab)}^{\kappa} - 1) H^{\kappa} (z_1)} e^{i(\phi_{(bc)}^{\kappa} - 1/2) H^{\kappa}(z_2)}e^{i(\phi_{(ca)}^{\kappa} - 1/2) H^{\kappa}(z_3)} e^{-\frac{i}{2} H^{\kappa}(w_1)}e^{-\frac{i}{2} H^{\kappa}(w_2)}\ket
\label{BigEq}\end{multline}
where $z_1$ has been fixed to $0$, and the angles $\phi^{\kappa}_{ab}$, $\phi^{\kappa}_{bc}$ and $\phi^{\kappa}_{ca}$ are external (hence $\phi^{\kappa}_{ab} + \phi^{\kappa}_{bc} + \phi^{\kappa}_{bc} = 2$ and $\sum_{\kappa=1}^3 \phi^{\kappa} = 2$). $C^{\alpha \beta}= i(\Gamma^1 \Gamma^2)^{\alpha \beta}$ is the charge conjugation operator. The above amplitude can be evaluated using the techniques outlined in \cite{Abel:2004ue, Abel:2005qn}. The perhaps unexpected form of the spin-field correlator is explained in appendix \ref{4d}. The most non-trivial part of the above is that involving the boundary-changing operators, which are dominated by worldsheet-instanton effects. We split $\partial X = \partial X_{qu} + \partial X_{cl}$, for which $\partial X_{qu}$ has boundary conditions such that all vertices have no displacements, whereas $\partial X_{cl}$ absorbs the displacements between the vertices. We have
\beq
\bra \partial \bar{X}_{qu} \prod_{i=1}^N \sigma_{\phi_i} \ket = 0
\eeq
and thus the amplitude is dominated by worldsheet instanton effects. To show the above, we consider
\begin{eqnarray}
\frac{\bra \partial \bar{X} (w) \prod_{i=1}^N \sigma_{\phi_i} (z_i) \ket}{ \bra \prod_{i=1}^N \sigma_{\phi_i} (z_i) \ket} &\sim& (w-z_i)^{-\phi_i} \qquad w \rightarrow z_i \nonumber \\
\frac{\bra \bar{\partial} \bar{X} (\bar{w}) \prod_{i=1}^N \sigma_{\phi_i} (z_i) \ket}{ \bra \prod_{i=1}^N \sigma_{\phi_i} (z_i) \ket} &\sim& (\bar{w}-\bar{z}_i)^{\phi_i-1} \qquad \bar{w} \rightarrow \bar{z}_i 
\end{eqnarray}
and construct a set of differentials satisfying the above local monodromies and periodicity of the worldsheet; these are the differentials given in \cite{Atick:Twist,Abel:2004ue, Abel:2005qn}. We then apply the global monodromy
\beq
\int_{\gamma_a} \d w \partial \bar{X} + \d \bar{w} \bar{\partial} \bar{X} = \bar{v}_a
\eeq
where $\gamma_a$ are a set of $N$ paths on the worldsheet and $v_a$ are the displacements between the vertices in the target space. There are $N$ independent differentials that comprise $\partial \bar{X}$ and $\bar{\partial} \bar{X}$, and so the global monodromies determine the coefficients by linear algebra. Since the paths are independent, the equations are non-degenerate and for $v_a = 0$ we must set all of the coefficients to zero, establishing the claim above. Defining the $N\times N$ matrix $W$ ($i^{\prime}$ runs through the set $\{z_1,z_2,...,z_{N-2}\}$ and $i^{\prime\prime}$ denotes the complementary set $\{z_{N-1},z_N\}$):
\begin{eqnarray}
W^{i^\prime}_a &=& \int_{\gamma_a} \d z\omega^{i^\prime} (z) \nonumber \\
W^{i^{\prime\prime}}_a &=& \int_{\gamma_a} \d \bar{z} \bar{\omega}^{i^{\prime\prime}} (\bar{z})
\end{eqnarray}
in terms of the $N$ cut differentials $\{\omega^{i^\prime},\omega^{i^{\prime\prime}}\}$, we can then write (after applying the doubling trick to relate $\partial X $ to $\bar{\partial} \bar{X}$)
\begin{eqnarray}
\bra \partial \bar{X} (x) \prod_{i=1}^N \sigma_{\phi_i} (z_i) \ket &=& -\bar{v}_a (\bar{W}^{-1})^{a}_{i^{\prime\prime}} \omega^{i^{\prime\prime}} (x) \bra \prod_{i=1}^N \sigma_{\phi_i} (z_i) \ket \nonumber \\
\bra \bar{\partial} \bar{X} (\bar{x}) \prod_{i=1}^N \sigma_{\phi_i} (z_i) \ket &=& -\bar{v}_a (\bar{W}^{-1})^{a}_{i^{\prime}} \bar{\omega}^{i^{\prime}} (-\bar{x}) \bra \prod_{i=1}^N \sigma_{\phi_i} (z_i) \ket .
\end{eqnarray}
The correlator is thus directly proportional to the displacements. Specialising now to the specific three-point function and using the prescription of \cite{Abel:2005qn} we have cycles $\{\gamma_a\} = \{A,B,C_2\}$ where $A$ and $B$ are the canonical cycles of the torus, and $C_2$ is the path passing between two vertices on the worldsheet. For this diagram, in each subtorus there is one brane parallel to the $E2$ brane, and this represents a periodic cycle on the worldsheet. The prescription for the amplitude requires that we permute the vertices cyclically so that the periodic cycle passes along the real axis; writing $\{a^{\kappa},b^{\kappa},c^{\kappa}\}$ for the cyclic permutation of branes $\{a,b,c\}$ such that brane $a^{\kappa}$ is parallel to the E2 brane in torus $\kappa$, we have
\begin{eqnarray}
v_A &=& \frac{1}{\sqrt{2}} n_A L_{a^{\kappa}} \nonumber \\
v_B &=& i\sqrt{2} (n_B \frac{4\pi^2 T_2^{\kappa}}{L_{a^{\kappa}}^{\kappa}} + y^{\kappa} ) \nonumber \\
v_{C_2} &=& \frac{1}{\sqrt{2}} e^{i\phi_{a^{\kappa}c^{\kappa}}^{\kappa}} ( n_C L_{c^{\kappa}}^{\kappa} + \Delta^{\kappa} )
\end{eqnarray}
where $\Delta^{\kappa}$ is the shortest distance between the target-space intersections $b^{\kappa}c^{\kappa}$ and $c^{\kappa}a^{\kappa}$ along the brane $c^{\kappa}$, and $y^{\kappa}$ is the distance between $a^{\kappa}$ and the E2-brane. The phase $e^{i\phi_{a^{\kappa}c^{\kappa}}^{\kappa}}$ in the last line appears due to the orientation of brane $a^{\kappa}$ relative to $c^{\kappa}$. 

The exponential of the worldsheet instanton action depends upon the same displacements,
and the amplitude can then be schematically written as 
\begin{multline}
\langle V^{0}_{\phi_{ab}} V^{1/2}_{\psi_{bc}} V^{1/2}_{\psi_{ca}} V^{-1/2}_{\theta} V^{-1/2}_{\theta}\rangle_{a,E2} = \phi_{(ab)} \psi_{(bc)\alpha} C^{\alpha \beta} \psi_{(ca)\beta} \big(\theta_1 \theta_2 - \theta_2 \theta_1\big) \\
\int \d t \int \d z_2 \d z_3 \ \sum_{i=1}^6 \prod_{\kappa=1}^3 \bigg( \sum_{n^{\kappa}_A,n^{\kappa}_B,n^{\kappa}_C} \frac{(f_i^{\kappa})^a \bar{v}_a^{\kappa}}{\sqrt{\ap}} e^{-\frac{S^{\kappa}}{2\pi\ap}} \bigg).
\end{multline}
Note that the action which appears here is the one-loop action as derived from the 
monodromy conditions, and depends on the integration variables.
The crucial part is that the functions $f_i$ arise from the different
permutations of applications of the picture-changing operators. Each
choice of $f_i$ corresponds to a different contribution that is
separately factorisable across the tori (and different from the
perturbative Yukawa term owing to the $\bar{v}_a^{\kappa}$
factors). Factorisability is in general lost upon performing the 
integral since there are no poles. However, since the functions are different, 
even if the integrals
were dominated by a particular region of the moduli space, we would have
Yukawa matrix corrections of the form
\beq
Y_{ab} \supset \sum_{i=1}^6 A^i_a B^i_b \, .
\eeq
This is a sum of six independent rank one matrices, giving a
rank three Yukawa matrix as advertised in the introduction. Note that
the correction terms are suppressed relative to the perturbative
superpotential by approximately the factor $e^{-S_{E2}^0}$; this
provides not only rank 3 couplings but an explanation for the
hierarchy in masses between the top quark and the others.
At the level of this analysis there is no obvious explanation 
for the hierarchies between the 1st and 2nd generation; this could 
yet arise from the non-factorisation of the worldsheet instanton contribution. 
We leave this issue for future work.

\appendix

\section{Partition Function for Massive N=2 Sectors}
\label{N2Sector}

In this appendix we evaluate the integral
\beq
Z(E2, D6_a) = N_a I_{E2,a}^{j} I_{E2,a}^{k} \int \frac{\d t}{t} \sum_{r^{i},s^{i}} \exp \bigg[ \frac{-8\pi^3 \alpha^{\prime} t}{(L_{E2}^{i})^2} |r^{i} + \frac{iT_2^{i} s^{i}}{\alpha^{\prime}} + \frac{i y_{a,E2} L_{E2}^{i}}{4\pi^2 \ap}|^2 \bigg] .
\eeq
First we must Poisson-resum the expression on $r$ and $s$ to obtain
\begin{eqnarray}
Z(E2, D6_a) &=& N_a I_{E2,a}^{j} I_{E2,a}^{k} \frac{(L_{E2}^i)^2}{T_2^i} \frac{1}{8\pi^2} \\
&& \times \int_0^{\infty} \d l \sum_{m,n} \exp  \bigg[-\frac{l (L_{E2}^i)^2}{8\pi} \bigg(\frac{1}{\ap} m^2 + \frac{\ap}{(T_2^i)^2} n^2 \bigg) 
  - 2i n \frac{y_{a,E2} L_{E2}^i}{2\pi T_2^i}\bigg] \, . \nonumber 
\end{eqnarray}
We then note the divergence for the piece $m,n=0$ as $l \rightarrow \infty$ 
(and note that the separation between the branes will regulate any $l\rightarrow 0$ 
infra-red divergence). 
This arises in the NS sector and is cancelled for consistent models according to the 
condition \cite{Lust:Gauge}
\beq
\sum_{a=\{a,a^{\prime}\}} \frac{(L_{E2}^i)^2}{T_2^i} N_a I_{E2,a}^{j} I_{E2,a}^{k} - 4 \sum_{\hat{k}} \frac{(L_{E2}^i)^2}{T_2^i} I_{E2,O6_{\hat{k}}}^{j} I_{E2,O6_{\hat{k}}}^{k} = 0.
\eeq
This is sufficient if there are no intersections between the $E2$-brane and the $D6$-branes of the model; if there are branes for which there are no parallel dimensions then the cancellation condition will include those. 

We now rescale the variables and perform the integral:
\beq
Z(E2, D6_a) = N_a I_{E2,a}^{j} I_{E2,a}^{k} \frac{1}{\pi} \sum_{m,n \ne 0}  \frac{\ap}{T_2^i} \frac{\exp \bigg[ - 2\pi in\frac{y_{a,E2} L_{E2}^i}{2\pi^2  
T_2^i} \bigg]}{|m + i\frac{(\ap)}{T^i_2}n|^2} .
\eeq
This can be recognised as appearing in Kronecker's second limit formula:
\begin{multline}
(2\Im(z))^s \sum_{(m,n) \ne (0,0)} \exp \bigg[ 2\pi i (m u + n v)\bigg] |m + n z |^{-2s} = \\
-4\pi \log | e^{-\pi u (v - u z)} \frac{\theta_1 (v- uz,z)}{\eta(z)} |  
 + O(s-1) 
 \end{multline}
and thus we obtain
\beq
Z(E2, D6_a) = 
- N_a I_{E2,a}^{j} I_{E2,a}^{k} \bigg[\log \left| \frac{\theta_1 (\frac{i y_{a, E2} L_{E2}^i}{2\pi^2 \ap}, 
\frac{i T_2^i}{\ap})}{\eta(\frac{iT_2^i}{\ap})} \right|^2 - \frac{y_{a, E2}^2}{2\pi^3 \ap} 
\frac{(L_{E2}^i)^2}{T_2^i}\bigg]\, .
\label{OneParallel}\eeq

\section{4d Spin Field Correlators}
\label{4d}

In the evaluation of annulus contributions to the superpotential
involving two fermionic fields and two supersymmetry modes, it is
necessary to evaluate the correlator of four left-handed spin fields
in four dimensions. In general, there may also be picture-changing
operators in the amplitude, although there are not for the particular
case in section \ref{YukCup}. The calculation is performed using the
techniques of \cite{Atick:1986rs}; the procedure is to construct a
complete set of Lorentz structures and determine their coefficients by
finding particular values for the spinor/Lorentz indices for which
only one structure is non-zero, and evaluating the correlator in those
cases. For the general case when there are two picture-changing
operators inserted on the non-compact directions for four
like-chirality spinors $\{u_1,u_2,u_3,u_4\}$, the amplitude is given
by
\begin{multline}
\bra \Psi^{\mu} (z) \Psi^{\nu} (w) \prod_{i=1}^4 (u_i)_{\alpha_i} \S^{\alpha_i} (z_i) \ket = -G(u_2 u_4) (u_3 C\Gamma^{\mu_3} \Gamma^{\mu_4} u_1) + J (u_1 u_2) (u_3 C\Gamma^{\mu_3} \Gamma^{\mu_4} u_4) \\
+H(u_3 u_2) (u_1 C\Gamma^{\mu_3} \Gamma^{\mu_4} u_4)  + 2B \eta^{\mu_3 \mu_4} (u_1 u_3) (u_2 u_4) + 2C  \eta^{\mu_3 \mu_4} (u_1 u_4) (u_2 u_3)
\label{Apsi4munu}\end{multline}
where $(u_1 u_2) \equiv (u_1)_{\alpha_1}C^{\alpha_1 \alpha_2} (u_2)_{\alpha_2}$, and the coefficient functions are given by
\begin{eqnarray}
G &=&  \bra \S^2 (z_1) \S^2 (z_2) \S^2 (z_3) \S^1 (z_4) \Psi^{0} (z) \Psi^{\bar{1}} (w) \ket \nonumber \\
J &=& \bra \S^1 (z_1) \S^2 (z_2) \S^2 (z_3) \S^2 (z_4) \Psi^{0} (z) \Psi^{\bar{1}} (w) \ket \nonumber \\
H &=&  \bra \S^2 (z_1) \S^2 (z_2) \S^1 (z_3) \S^2 (z_4) \Psi^{0} (z) \Psi^{\bar{1}} (w) \ket \nonumber \\
B &=& \bra \S^2 (z_1) \S^1 (z_2) \S^1 (z_3) \S^2 (z_4) \Psi^{0} (z) \Psi^{\bar{0}} (w) \ket \nonumber \\
C &=& \bra \S^1 (z_1) \S^2 (z_2) \S^1 (z_3) \S^2 (z_4) \Psi^{0} (z) \Psi^{\bar{0}} (w) \ket.
\end{eqnarray}
Note that we have written the functions in terms of gamma matrices rather than the standard Weyl-notation matrices $\sigma^{\mu \nu}$ since the amplitude with PCOs is summed over momenta - and it is then possible to cancel many terms via the on-shell conditions. We sacrifice obvious antisymmetry on the inserted operators, but it is straightforward to show that it is still antisymmetric on exchange of $\psi^{\mu_3}(z)$ and $\psi^{\mu_4}(w)$. To demonstrate that the above is a complete set, we require the standard Fierz identities, but we also require corresponding identities among the products of Jacobi Theta functions. In particular, we require
\begin{multline}
\th1.3 \th2.4 \theta_1 (z - z_1) \theta_1( z- z_3) \theta_1 (w - z_2) \theta_1 (w - z_4) \\
\theta_{\nu} (\frac{z_2 + z_4 - z_1 - z_3}{2}) \theta_{\nu} (\frac{z_1 + z_3 - z_2 - z_4}{2} + z - w) \\
- \th1.2 \th1.3 \th2.3 \theta_1(z - z_4) \theta_1(w - z_4) \theta_1 (z - w) \\
 \theta_{\nu} (\frac{z_4 - z_1 - z_3 - z_2}{2} + z) \theta_{\nu} (\frac{z_3 + z_2 - z_4 + z_1}{2} - w) \\
- \th1.4 \th2.3 \theta_1 (z- z_2) \theta_1( z- z_3) \theta_1 (w - z_1) \theta_1 (w - z_4) \\
\theta_{\nu} (\frac{z_1 + z_4 - z_2 - z_3}{2}) \theta_{\nu} (\frac{z_2 + z_3 - z_1 - z_4}{2} + z - w) \\
- \th1.2 \th3.4 \theta_1 (z- z_1) \theta_1( z- z_2) \theta_1 (w - z_3) \theta_1 (w - z_4) \\
\theta_{\nu} (\frac{z_3 + z_4 - z_1 - z_2}{2}) \theta_{\nu} (\frac{z_1 + z_2 - z_3 - z_4}{2} + z - w) \\
= 0.
\label{Identity}\end{multline}
To prove this, it is easy to check that the periodicities of the terms are the same, and when one of the functions is zero, the remaining three sum to zero. Thus we can write any one of the functions as a constant multliple of the other three; since we can do this for any function the constant must be $-1$, so the identity holds in general.

The reader may then substitute $\theta_{\alpha_3}, \theta_{\alpha_4}$ for $(u_{3})_{\alpha_{3}}, (u_4)_{\alpha_4}$. This results in many simplifications, because we need only keep structures involving $\theta_1 \theta_2 = -\theta_2 \theta_1$. 
However, to obtain the amplitude without PCO insertions, we can use the OPE of the $\Psi$ fields in the above. Alternatively, we write the amplitude as
\beq
\bra \prod_{i=1}^4 (u_i)_{\alpha_i} \S^{\alpha_i} (z_i) \ket = A_1 (u_1 u_3) (u_2 u_4) + A_2 (u_1 u_4) (u_2 u_3)
\eeq
where 
\begin{eqnarray}
A_1 &=& \bra \S^1 (z_1) \S^2 (z_2) \S^2 (z_3) \S^1 (z_4)\ket \nonumber \\
A_2 &=& \bra \S^1 (z_1) \S^2 (z_2) \S^1 (z_3) \S^2 (z_4)\ket.
\end{eqnarray}
Here we have used 
\begin{eqnarray}
(u_1 u_3) (u_2 u_4) - (u_1 u_4) (u_2 u_3) &=& (u_1 u_2) (u_3 u_4) \nonumber \\
A_1 - A_2 &=& \bra \S^1 (z_1) \S^1 (z_2) \S^2 (z_3) \S^2 (z_4)\ket.
\end{eqnarray}
Substitution of $\theta_{\alpha_3}, \theta_{\alpha_4}$ for $(u_{3})_{\alpha_{3}}, (u_4)_{\alpha_4}$ and disregarding terms proportional to $\theta_1^2$ and $\theta_2^2$ results in the expression given in equation \ref{BigEq}.


\section{Spin-Structure Summation}
\label{SpinSum}

It is possible to compute the spin-structure summation for the expression \ref{BigEq}, since the non-compact spin struture is partially cancelled by the spin-dependent part of the superconformal ghost amplitude. The result is
\begin{multline}
\langle V^{0}_{\phi_{ab}} V^{1/2}_{\psi_{bc}} V^{1/2}_{\psi_{ca}} V^{-1/2}_{\theta} V^{-1/2}_{\theta}\rangle_{a,E2} = \phi_{(ab)} \psi_{(bc)\alpha} C^{\alpha \beta} \psi_{(ca)\beta}\  \theta_1 \theta_2  \\
\int \d t \prod_{i=1}^3 \int_0^{it} \d z_i \  f(z_2-z_3,t) \lim_{x_1 \rightarrow z_1} \lim_{x_2\rightarrow z_2} \lim_{x_3 \rightarrow z_3}  (x_1-z_1) (x_2-z_2)^{1/2} (x_3-z_3)^{1/2}\\
\theta_1 (z_2 - z_3)^{1/4} \bra \prod_{i=1}^3 e^{i k_i \cdot X (z_i)} \ket  \sum_{\{y_1,y_2,y_3\} = P(x_1,x_2,x_3)} \prod_{\kappa=1}^3 \sqrt{\frac{2}{\ap}}\langle \partial \bar{X}^{\kappa} (y_\kappa) \sigma_{\phi^{\kappa}_{ab}} (z_1) \sigma_{\phi^{\kappa}_{bc}} (z_2)\sigma_{\phi^{\kappa}_{ca}}(z_3) \ket \\
\bra e^{i H^{\kappa} (y_{\kappa})} e^{i(\phi_{(ab)}^{\kappa} - 1) H^{\kappa} (z_1)} e^{i(\phi_{(bc)}^{\kappa} - 1/2) H^{\kappa}(z_2)}e^{i(\phi_{(ca)}^{\kappa} - 1/2) H^{\kappa}(z_3)}\ket_{(\nu\ \mathrm{indep.})} \\ 
\theta_{1} ((\phi_{(ab)}^{\kappa} -1)z_1 + (\phi_{(bc)}^{\kappa} -1)z_2 +(\phi_{(ca)}^{\kappa} -1)z_3 + y_{\kappa}) 
\end{multline}
where
\begin{equation}
\bra \prod_{i=1}^N e^{i a_i H (z_i)} \ket_{(\nu \ \mathrm{indep})} \equiv \prod_{i < j} \theta_1 (z_i - z_j)^{a_i a_j}
\end{equation}
and
\begin{equation}
f (\delta,t) \equiv 2\exp[\frac{\pi t}{2}]\prod_{j=1}^2 \int_{0}^{it} \d w_j^{\prime} \frac{\theta_1 (w_1^{\prime} - w_2^{\prime} + \delta) \theta_4^2 (w_1^{\prime} + w_2^{\prime})}{\theta_1 (w_1^{\prime} + w_2^{\prime})\theta_2 (w_1^{\prime})\theta_2 (w_2^{\prime}) \theta_2 (w_1^{\prime} - \delta )\theta_2 (w_2^{\prime}+ \delta)}
\end{equation}
encodes all of the dependence on the position of the $V_{\theta}$ insertions. It is an odd function of $\delta$, and hence the amplitude does not have a pole at $z_2 = z_3$, as required. 

\bibliography{e2inst}

\end{document}